\documentclass[prb,aps,twocolumn,showpacs,nofootinbib]{revtex4}
\usepackage{graphicx} 
\usepackage{amsmath}
\usepackage{sistyle}
\newcommand{\mbR}{\mathbf{R}}
\newcommand{\phase}{\mathrm{\sc P}}
\newcommand{\mbr}{\mathbf{r}}
\newcommand{\mbk}{\mathbf{k}}
\newcommand{\mbQ}{\mathbf{Q}}
\newcommand{\mbE}{\mathbf{E}}
\newcommand{\mbp}{\mathbf{p}}
\newcommand{\mbg}{\mathbf{g}}
\newcommand{\mhy}{\mathbf{\hat y}}
\newcommand{\mhx}{\mathbf{\hat x}}
\newcommand{\mhz}{\mathbf{\hat z}}

\begin{document}
%
\title{Diffractive arrays of gold nanoparticles near an interface: critical role of the substrate}
\author{Baptiste~Augui\'e,$^{1,2}$ Xes\'us~M.~Benda\~{n}a,$^1$ William~L.~Barnes,$^3$ \\ and F.~Javier~Garc\'{\i}a de Abajo$^1$}
\email{J.G.deAbajo@csic.es}
\affiliation{$^1$Instituto de \'Optica - CSIC and Unidad Asociada CSIC-Universidade de Vigo, Serrano 121, 28006 Madrid, Spain\\ $^2$Departamento de Qu\'{\i}mica F\'{\i}sica and Unidad Asociada CSIC-Universidade de Vigo, 36310 Vigo, Spain\\ $^3$School of Physics, University of Exeter, Stocker Road, Exeter, Devon, EX4 4QL, UK}

\date{\today}

\begin{abstract}
The optical properties of periodic arrays of plasmonic nanoantennas are strongly affected by coherent multiple scattering in the plane of the array, which leads to sharp spectral resonances in both transmission and reflection when the wavelength is commensurate with the period. We demonstrate that the presence of a substrate (i.e., an asymmetric refractive-index environment) can inhibit long-range coupling between the particles and suppress lattice resonances, in agreement with recent experimental results. We find the substrate-to-superstrate index contrast and the distance between the array and the interface to be critical parameters determining the strength of diffractive coupling. Our rigorous electromagnetic simulations are well reproduced by a simple analytical model. These findings are important in the design of periodic structures and in the assessment of their optical resonances for potential use in sensing and other photonic technologies.
\end{abstract}
\maketitle
\section{Introduction}
Scattering of light by periodic metallic structures has been well studied for over a century in the context of diffraction gratings.\cite{R1907-2,W1902,F1941} The subject has recently received renewed attention\cite{ZS04,MS07} with the prediction\cite{M05-2,HZS05} and experimental observation\cite{LCK01,KSG08,AB08,CSY08,VGG09} of interesting optical phenomena that result from the interaction between the geometrical resonance associated with light diffraction and the excitation of localized surface-plasmon resonances in metallic nanoparticles, which play the role of plasmonic nanoantennas.\cite{N07_2} In addition to the interesting physics revealed in such systems, a number of applications have been proposed, including nanoscale energy transport,\cite{MKA03,paper148} sensing,\cite{EQB04, KSK10} and modifying spontaneous emission,\cite{VGG09-2} which rely on the improved quality factor resulting from the reduction in radiative damping of the array as compared to localized plasmons excited in isolated particles. Recent advances in the control of the angular emission from quantum dots are also based on diffractive coupling of antenna elements.\cite{CVT10}

Two-dimensional arrays of nanoantennas can be produced with good fabrication control by techniques such as electron-beam lithography,\cite{LCK01,KSG08,AB08,CSY08,AB09} contact printing,\cite{VGG09} and colloidal chemistry.\cite{GPM08} These samples are most commonly manufactured on a substrate of high refractive index compared to the upper medium (typically air, or water for biosensing applications). It has been suggested from experimental data\cite{AB08} and theoretical modeling\cite{paper186} that such an asymmetric configuration is incompatible with the existence of delocalized surface modes, and therefore prevents the observation of efficient narrowing of the localized surface-plasmon lineshapes. This is in contrast to other studies of fluorescence in particle arrays.\cite{VGG09-2}

Here, we elucidate the conditions under which strong diffractive coupling may occur in the asymmetric configuration. In particular, we examine the reflectivity of gold nanoantenna arrays in close proximity to a substrate. A sharp transition is observed when varying the particle size above a certain threshold, leading to the sudden emergence of lattice resonances, and establishing a clear difference in the behavior of lithographically patterned nanoantenna arrays depending on the thickness of the metal layer. Likewise, we find the refractive-index contrast and the array-substrate separation to be critical parameters.
\section{Modeling particle arrays near a substrate}
\label{sec:theory}
\begin{figure*}[!ht]
\centering \includegraphics[width=\textwidth]{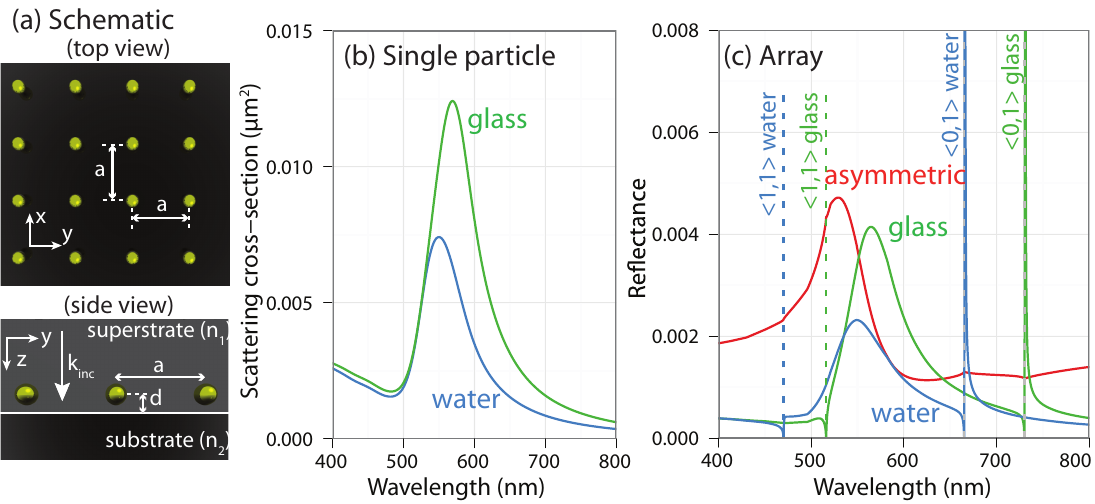}
\caption{(a) Schematic representation of the system under study. An infinite periodic two-dimensional square array of gold nanoparticles is situated in a semi-infinite, homogeneous medium of refractive index $n_1$ at a distance $d$ above a substrate of refractive index $n_2$. The center-to-center separation between particles is $a$.  Light is incident from the superstrate medium with a direction normal to the plane of the array and polarization along a lattice vector ($\hat x$).  (b) Elastic scattering cross-section of an individual gold sphere immersed in water ($n=1.33$) and in glass ($n=1.46$), as obtained from Mie theory. (c) Same as (b) for a particle array of period $a=\SI{500}{nm}$ rather than an individual particle, also including the asymmetric configuration in which the spheres are in water with their surfaces located at a distance of \SI{1}{nm} from a glass substrate. Vertical dashed lines indicate the position of the \texttt{<0,1>} and \texttt{<1,1>} diffraction conditions in the different media. All gold nanospheres are taken to have a radius of \SI{35}{nm}.} \label{fig:schematic}
\end{figure*}

The system under study is schematically depicted in Fig.~\ref{fig:schematic}a. It consists of an infinite square array of gold spheres placed in water close to a glass substrate and illuminated under normal incidence. For the particle dimensions under consideration, the optical response of each individual sphere shows a characteristic dipole plasmon resonance, as shown in Fig.~1b for two different host materials. The interaction between the particles in the array leads to additional, sharper features that correlate with the diffracted orders of the lattice, as shown for some specific geometrical parameters in Fig.~1c. In particular, for a homogeneous environment, the \texttt{<0,1>} diffraction feature occurs when the wavelength in the medium is close to the period, $\lambda/n=a$. This condition is obviously dependent on the refractive index $n$ (Fig.~1c, blue and green curves). Strikingly, the diffraction feature disappears when the array is placed in an asymmetric environment, close to an interface (Fig.~1c, red curve). Interestingly, the reflectance maximum associated with the dipolar plasmon shifts to the blue in this configuration as a result of interference between light scattered from the particles and the wave reflected at the planar interface (the specularly reflected beam at the planar interface interferes with the beam scattered from the spheres, which undergoes large phase shifts as the wavelength sweeps across the plasmon resonance, thus distorting the lineshape, unlike what happens in non-specular reflectance, for which such interference is not present).

We set out to explain the origin of the suppression of diffraction close to the interface. The results of Fig.~1c are obtained from rigorous multiple-scattering numerical solution of Maxwell's equations,\cite{SKM92} with the gold described by a tabulated measured dielectric function\cite{JC1972} and the rest of the materials having wavelength-independent refractive index. However, this method of solution does not provide insight into the origin of the effect, which we explore next using a simple analytical model that allows us to determine the most relevant parameters involved in the phenomenon of inhibition of collective resonances in the arrays. A variety of semi-analytical and numerical techniques have been developed in this context.\cite{YYK74,MWL1985,J01,SM91,SH02,S09,S10,VFN10} For the geometrical parameters under consideration, it is reasonable to represent the particles as induced point dipoles, for which we obtain the polarizability $\alpha$ from the first Mie coefficient.\cite{KTE03}

In the array, the dipole moment $\mbp^j$ at each particle site $j$ satisfies the self-consistent coupled-dipoles equation
\begin{equation}
\mbp^j=\alpha\left[\mbE^\text{ext}(\mbR_j)+\sum_{j'\neq j}\mathcal{G}^0(\mbR_j-\mbR_{j'})\mbp^{j'}\right],
\label{eq1}
\end{equation}
where $\mbE^\text{ext}$ is the external electric field and the sum gives the field induced by other particles $j'$ at the lattice position $\mbR_j$. Here, $\mathcal{G}^0({\bf r})=(k_1^2+\nabla\nabla)\exp(ik_1r)/r$ is a Green tensor that yields the electric field produced by a dipole in the surrounding homogeneous medium of refractive-index $n_1$ and light wave vector $k_1=kn_1$, where $k$ is the free-space wave vector.

When the wavelength in the medium $2\pi/k_1$ is commensurate with the period of the array $a$, scattering of the incident light by the array produces diffracted beams. Only some of these beams are propagating, but the remaining evanescent waves play an important role. For wavelengths above the \texttt{<0,1>} threshold in the spectra of Fig.~1c, only the specular beam is propagating, but four additional beams $<\pm1,\pm1>$ become propagating at shorter wavelengths. These beams are evanescent above the offset wavelength for diffraction, but in the presence of a substrate they may undergo multiple reflections in the cavity formed between the array and the interface, thereby contributing to the reflectivity of the combined system. Thus our approach is to model the inhomogeneous environment using a Green tensor based upon the homogeneous environment and to add the effect of the substrate through (multiple) reflections from the substrate.

The influence of the substrate on the beams is further illustrated in Fig.~2a, which presents dispersion diagrams in the form of reflectance as a function of the total and parallel light wave vectors, $k_1$ and $k_\parallel$, respectively. The curves in Figure\ 1c correspond to cuts of the color plots of Figure\ 2a along the vertical axis. For homogeneous environments (left and central plots), diffraction features are clearly visible near $k_1=2\pi/a$. However, these features disappear in the presence of a substrate (right plot). Figure\ 2b shows that the \texttt{<0,1>} feature for normal incidence can receive contributions from diffracted beams with $k_\parallel=2\pi/a$ via momentum exchange with the lattice. Therefore, these diffraction beams have to be incorporated into our model, and we show below that the substrate enhances their effect to the point of suppressing diffraction.

\begin{figure}[!ht]
\centering
\centering \includegraphics[width=\columnwidth]{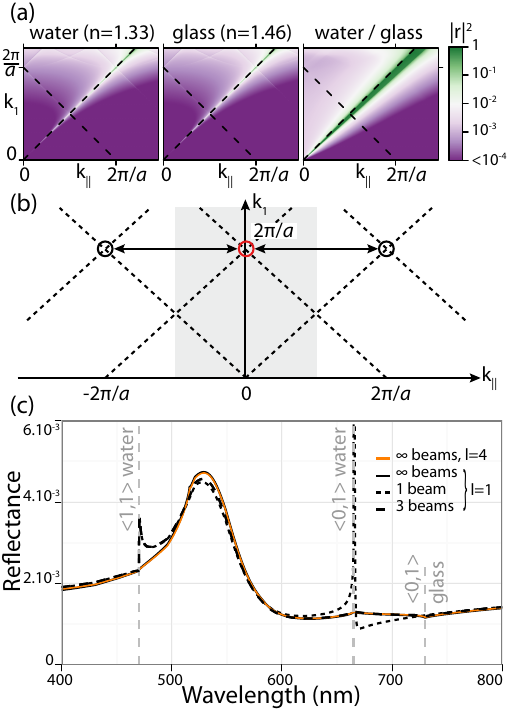}
\caption{(a) Dispersion diagrams showing the reflectance of square arrays of \SI{35}{nm}-radius gold particles immersed in water (left), in glass (center), and in water at a surface-to-surface distance of \SI{1}{nm} from a water-glass interface (right, see Fig.~1a). Here, $k_1$ is the wave vector in the host medium, $k_\parallel$ is the parallel wave vector along a principal lattice direction, and $a=\SI{500}{nm}$ is the lattice period. The incident polarization is TE. (b) Scheme showing the first Brillouin zone (shaded area) and its vicinity. The dashed lines (light line and diffracted light line) correspond to the condition of grazing low-order diffracted beams. The circles signal the three beams contributing to the \texttt{<0,1>} diffraction spectral anomaly under normal incidence via $(0,\pm 2\pi/a)$ wave-vector transfers from the lattice (double-arrowed horizontal lines). (c) Calculated normal-incidence specular reflectance of the square array in the asymmetric environment considered in (a). Rigorous numerical solutions of Maxwell's equations including multipoles with orbital angular momentum up to $l=4$ are compared with a simplified solution using only dipoles ($l=1$). These results, which require a large number of diffracted beams $\sim 20$ to get convergence, are compared with our analytical model for dipoles [Eq.~(\ref{eq:R11})] using only one beam (zero-order, red circle in (b)) or three beams (all circles in (b)).} \label{fig:gmax}
\end{figure}

It should be noted that diffraction originates in long-distance interaction between particles, which are polarized along the direction of the external field, $\mhx$. Now, the dipolar field dies off faster than $1/r$ with the distance along $\mhx$, and therefore, the $<\pm1,0>$ beams are not expected to play a significant role. Thus, in our minimal model we only include three beams with parallel wave vectors $\mbg_0=(0,0)$ (specular beam) and $\mbg_{\pm1}=\pm(2\pi/a)\mhy$ (diffracted beams).

For an incident TE-polarized beam of parallel wave vector $\mbg_m$ with unit electric field along $\mhx$, the induced dipoles have the form $\mbp^j=p_m\exp(i\mbg_m\cdot\mbR_j)\,\mhx$, which upon insertion into Eq.~(\ref{eq1}) yields
\begin{equation}
p_m=\frac{1}{1/\alpha - G_m},
\label{eqp}
\end{equation}
where $G_m=\sum_{j\neq0}\mathcal{G}^0\left(\mbR_j\right)\exp(-i\mbg_m\cdot\mbR_j)$ is a lattice sum representing the collective interaction between dipoles (notice that the $\mbR_0=0$ site is excluded from the sum).\cite{paper130} Both $p_m=p$ and $G_m=G$ are independent of $m$ for the three $\mbg_m$ beams under consideration. We show 
\begin{eqnarray}
G=\sum_{j\neq0}\mathcal{G}^0\left(\mbR_j\right)
\label{eqG}
\end{eqnarray}
in the Appendix.

The reflection coefficients of the array can be obtained upon examination of the scattered field. In particular, the $x$ component reduces to
\begin{equation}
E_x^\text{scat}=\sum_j\left(k_1^2+\frac{\partial^2}{\partial x^2}\right)\frac{e^{ik_1\mbr-\mbR_j|}}{|\mbr-\mbR_j|}\;p.
\label{eq2}
\end{equation}
Now it is useful to expand the spherical waves of this expression in parallel wave vector space as
\begin{equation}
\frac{e^{ik_1r}}{r} = \int\frac{d^2\mbQ}{(2\pi)^2} \frac{2\pi i} {k_z} e^{i\mbQ\cdot \mbr + ik_z |z|},
\label{eq3}
\end{equation}
where $k_z=\sqrt{k_1^2-Q^2}$ is the normal wave vector component. Inserting Eq.~(\ref{eq3}) into Eq.~(\ref{eq2}) and using the relation $\sum_j\exp(i\mbQ\cdot\mbR_j)=(4\pi^2/a^2)\sum_{\mbg}\delta (\mbQ-\mbg)$, we find
\begin{equation}
E_x^\text{scat} = \sum_{\mbg}\frac{2\pi i} {a^2k_z} e^{i\mbk_\pm\cdot\mbr}\left(k_1^2-g_x^2\right) p,
\nonumber
\end{equation}
where $\mbg$ runs over reciprocal lattice vectors and $\mbk_\pm = \mbg \pm k_z\mhz$. Noticing that $g_x=0$ for the three beams included in our model, the scattered field can be approximated by
\begin{equation}
E_x^\text{scat}\approx\sum_m\frac{2\pi ik_1^2 } {a^2k_z}e^{i\mbk_\pm\cdot\mbr}p,
\label{eq4}
\end{equation}
in which both $\mbk_\pm$ and $k_z$ depend on $m$ through the lattice vector $\mbg_m$.
The reflection coefficient of the array $r^a$ can be expressed as a $3\times3$ matrix with coefficients $r^a_{jj'}$ relating the incident beam $j'$ to reflected beams $j$. Using Eqs.~(\ref{eqp}) and (\ref{eq4}), we find
\begin{equation}
r^a_{jj'}=\frac{2\pi i k_1^2}{a^2k_{zj}} \cdot \frac{1}{1/\alpha - G},
\label{eq:ra}
\end{equation}
where $k_{zj}=\sqrt{k_1^2-(2\pi m/a)^2}$.

The incident plane wave and the beams diffracted by the array are specularly reflected at the interface with the substrate, for which the coefficients of the reflection matrix $r^s$ are obtained from Fresnel's formula for TE polarization:
\begin{equation}
r_{jj'}^s = \delta_{jj'} \frac{k_{zj} - k'_{z_j}}{k_{zj} + k'_{zj}},
\nonumber
\end{equation}
where $k'_{zj}=\sqrt{k_2^2-(2\pi m/a)^2}$. Multi-layered substrates can be straight-forwardly included in this analysis by introducing a suitable reflection coefficient. In particular, the case of substrates supporting surface modes may introduce new spectral features~\cite{HO65}; this is however beyond the scope of our present study.

The composite array-substrate system forms an optical cavity, for which the total reflectivity must incorporate the effect of multiple internal reflections. A Fabry-Perot-type of analysis yields the combined reflection matrix
\begin{equation}
r = r^a + \frac{t^a\phase\ r^s\phase}{1- r^a\phase\ r^s\phase} t^a,
\label{eq6}
\end{equation}
where $\phase_{jj'}=\delta_{jj'}\exp{\left[ik_{zj}d\right]}$ describes plane-wave propagation across the cavity and the transmission matrix of the array satisfies $t^a_{jj'}=\delta_{jj'}+r^a_{jj'}$. Finally, Eq.~(\ref{eq6}) gives an analytical expression for the specular reflectivity coefficient of the zero-order beam,
\begin{equation}
r_{00}^\text{tot} = \frac{r^s_{00}e^{2ikd}\left[2  e^{2ik_{z1}d} r^a_{10} r^s_{01} - 2 r^a_{00} - 1\right] -r^a_{00} }{2 e^{2ik_{z1}d} r^a_{01} r^s_{11} + e^{2ikd} r^a_{00} r^s_{00} - 1}.
\label{eq:R11}
\end{equation}
This expression reveals an intricate dependence of the reflectance of the composite system on the various physical parameters. The phase factors $~\exp(2ik_{zj}d)$ yield a periodic modulation in regions far from intrinsic or lattice resonances (Fabry-Perot effect). The poles in Eq.~(\ref{eq:R11}) do not trivially expose the position of the modes supported by the structure because the numerator of Eq.~(\ref{eq:R11}) may compensate for a possible divergence. Lattice resonances may occur through two different sources. First, the coefficients $r^a$ contain a factor of the form $(1/\alpha - G)^{-1}$, which is responsible for the lattice resonance in a self-standing array near the diffraction edge. Second, the reflectivity coefficient $r^a_{01}$ for the grazing diffractive orders has a factor $k_{z}^{-1}=1/\sqrt{k_1-\frac{2\pi}{a}}$, which diverges right at the diffraction edge. It is precisely the interplay between the divergent terms in the numerator and the denominator of Eq.~(\ref{eq:R11}) that is responsible for the cancellation of the diffractive coupling in an asymmetric configuration.

This is clearly illustrated in Fig.~2c, in which we assess the accuracy of this model in describing the optical properties of an array of small gold spheres near a substrate. The  reflectance predicted by Eq.~(\ref{eq:R11}) (broken curves) is compared to a rigorous calculation based upon a multiple-scattering formalism described elsewhere\cite{SM91} (solid curves). The results from the analytical model are nearly identical to those of the full calculation near the \texttt{<0,1>} diffraction edge, and both predict a featureless spectrum in that region. In contrast, a simplified version of the model accounting for only the specular beam [dotted curve; Eq.~(\ref{eq6}) reduces to a scalar equation for the zero-order beam] shows a pronounced \texttt{<0,1>} feature. The conclusion is clear: diffracted beams cancel the long-range coupling between particles in the presence of a substrate. Physically, this cancellation occurs because of the interference between the reflected grazing diffracted orders and the direct dipolar coupling in the superstrate medium.

The accuracy of the three-beams model described by Eq.~(\ref{eq:R11}) is good near the \texttt{<0,1>} diffraction edge. However, at shorter wavelengths, and in particular around the \texttt{<1,1>} diffraction edge, both the zero-order beam (dotted curve) and three-beams model (dashed curve) predict a non-existent spectral peak. This feature disappears when more beams are incorporated into the analytical model.

Incidentally, we have also checked the validity of the dipolar approximation by including higher-order multipoles in the rigorous calculation (up to $l=4$). The results are nearly indistinguishable regardless the number of multipoles present in the full calculation, as expected for spheres that are much smaller than the wavelength.

In what follows, we investigate the influence of several physical parameters on the strength of the diffractive peak using the analytical model of Eq.~(\ref{eq:R11}).
\section{Results and discussion}
\label{sec:results}
\begin{figure}[!ht]
\centering
\centering \includegraphics[width=\columnwidth]{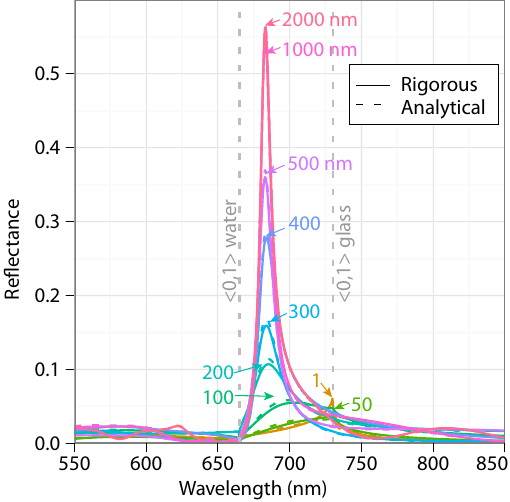}
\caption{Calculated normal-incidence specular reflectance of a square array of 60-nm-radius gold spheres immersed in water and placed above a glass substrate (see Fig.~1a). The lattice period is 500\,nm. The distance from the bottom of the spheres to the substrate is varied from \SI{1}{nm} to \SI{2}{\mu m} (see labels). Dashed curves: analytical model of Eq.~(\ref{eq:R11}). Solid curves: fully converged multiple-scattering numerical solution.}
\label{fig:distance}
\end{figure}

In Fig.~\ref{fig:distance} we investigate the dependence of the reflectance on the distance between the array and the substrate. The figure clearly shows the transition between the absence of diffraction when the particles are touching the interface and the sharp diffraction peak obtained at large distances, corresponding to the limit of a homogeneous environment. 
As the array-substrate distance is reduced, this mode decreases in strength and begins to red shift for separations below 200\,nm. Eventually, the mode disappears and a new, weaker peak is observed at the diffraction condition for the \emph{substrate} medium.


\begin{figure*}[!ht]
\centering
\centering \includegraphics[width=\textwidth]{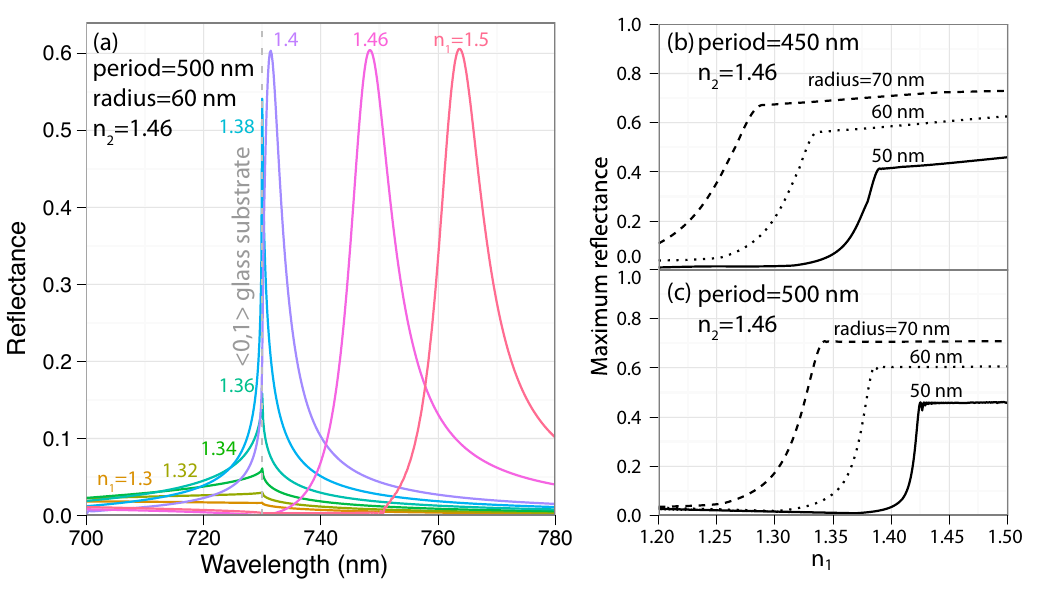}
\caption{(a) Calculated normal-incidence specular reflectance of a square array of \SI{60}{nm}-radius gold spheres and period equal to \SI{500}{nm} immersed in a fluid and placed at a surface-to-surface distance of \SI{1}{nm} from a glass substrate (see Fig.~1a). The refractive index of the fluid is varied in the range $n_1=1.2-1.5$ around the glass index $n_2=1.46$. (b,c) Maximum of the reflectance spectra as a function of $n_1$ for two values of the array period and three different sphere radii (see text insets). All spectra are calculated with the analytical model of Eq.~(\ref{eq:R11}).}
\label{fig:medium}
\end{figure*}


A similar transition occurs when the index of the substrate is made to match that of the superstrate. In Fig.~4a we show reflectance spectra for a range of index differences.  As the index asymmetry is reduced, the mode strength increases dramatically after the asymmetry is below a threshold value. This is clearly observed in the evolution of the peak maximum with index contrast (Fig.~4b,c). The threshold for diffraction inhibition seems to depend on both the period of the array and the size of the particles. In general, this threshold occurs at larger contrast when the particles are bigger or the period smaller, and the transition is smoother for smaller period. Actually, particles of larger size relative to the period deviate more from the ideal situation of small particles situated close to the interface, thus involving significant phase factors $\phase$ [see Eq.~(\ref{eq6})] that make the diffraction feature more robust against index contrast.
A less intuitive result is that the array period does not substantially affect the maximum reflectance for a symmetric medium, although the fractional occupancy of the spheres is inversely proportional to the period squared. This mode is associated with diffraction, and the reflectivity of the system is therefore strongly affected by coherent multiple-scattering, so that a simple geometric scaling rule is no longer applicable.

\begin{figure}[!ht]
\centering
\centering \includegraphics[width=\columnwidth]{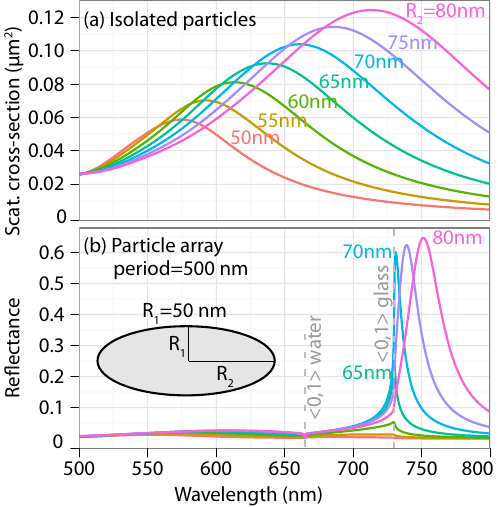}
\caption{(a) Scattering cross-section of gold prolate ellipsoids with increasing long-axis radius $R_2$ (see labels) and short-axis $R_1=\SI{50}{nm}$. The polarizability is derived from Ref.~\onlinecite{KTE03}. (b) Reflectance of ellipsoid arrays calculated from the analytical model of Eq.~(\ref{eq:R11}), using the polarizabilities of (a). The ellipsoids are immersed in water at a surface-to-surface distance of \SI{1}{nm} from a glass substrate (see Fig.~1a). The incident electric field is along the long axes of the particles ($R_2$), which are in turn oriented along a principal lattice direction. The diffraction edges associated with the water and glass media are indicated as vertical dashed lines.}
\label{fig:kuwata}
\end{figure}

Because spheres of increasing radius need to be placed further away from the substrate, the transition of Fig.~\ref{fig:medium} might perhaps be attributed to the effect discussed in Fig.~\ref{fig:distance} (the changing distance between the substrate and the sphere centers as the particle radius increases). To further discriminate between the two effects---the particle-centers separation from the substrate and the polarizability of the particles)---, we explore next arrays formed by elongated ellipsoids of increasing aspect ratio, showing that an increase in polarizability drives a transition between suppression and emergence of diffraction features.
Within the framework of the dipolar approximation used in our analytical model, we adopt the polarizability prescription developed by Kuwata-Gonokami \emph{et al.},\cite{KTE03} which provides a good approximation for subwavelength particles of moderate aspect ratio. Fig.~\ref{fig:kuwata}a presents results of this model for prolate gold ellipsoids immersed in water. The incident field is polarized along the long-axis of the particles. We have performed rigorous T-matrix calculations\cite{DWE06} to verify the accuracy of the model for the particles considered here. Elongating the long axis of the particles enables us to tune the polarizability without altering the distance between the particles and the substrate. The scattering cross-section exhibits a localized surface plasmon resonance that is red-shifted and stronger as the aspect ratio increases.

Using the polarizability of these particles in our analytical model, we obtain the results shown in Fig.~\ref{fig:kuwata}b for an array supported by a glass substrate. A transition very similar to that of Fig.~\ref{fig:medium}a is observed as the polarizability of the particles increases. The diffractive coupling in the superstrate medium is suppressed by the presence of the substrate for the \SI{50}{nm} spheres, but as the long-axis of the particles is increased a new peak gains strength at wavelengths above the diffraction condition in the substrate.

\begin{figure}[!ht]
\centering \includegraphics[width=\columnwidth]{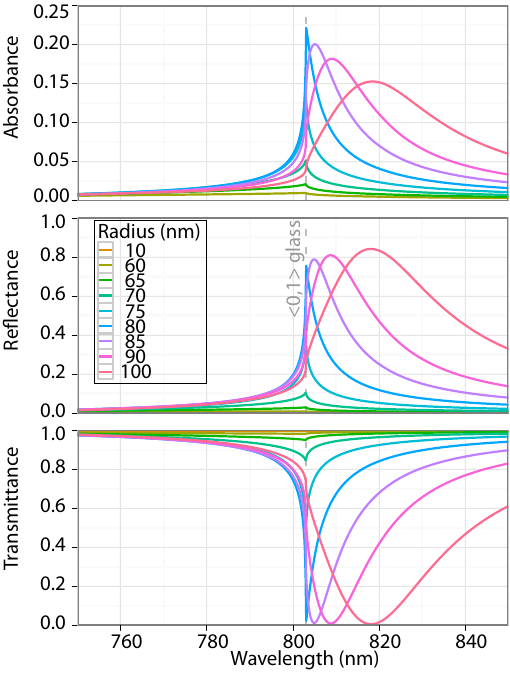}
\caption{Calculated zero-order transmittance, reflectance, and absorbance spectra for a square array of gold spheres with increasing radius as indicated in the legend (dielectric function from Johnson and Christy\cite{JC1972}). The array has pitch \SI{550}{nm} and is placed inside water (refractive index \num{1.33}). The sphere surfaces are separated \SI{1}{nm} from a glass substrate (refractive index \num{1.46}).}
\label{fig:transition-r}
\end{figure}

These results are qualitatively similar to those obtained for spheres of increasing radius. Fig.~\ref{fig:transition-r} represents the transmittance, reflectance, and absorbance of gold-sphere arrays immersed in water and supported on glass for various particle sizes and fixed pitch of the array. The vertical dashed line represents the onset of the \texttt{<0,1>} order diffraction. The arrays show an absorption and reflection feature for a wavelength slightly to the right of the onset when the particle radius exceeds a value $\sim \SI{60}{nm}$. This is the result of a lattice resonance involving the collective interaction of the spheres close to the condition for which the noted order of diffraction becomes grazing.\cite{R1907} The resonance is increasingly broadened and redshifted as the particle size increases. This effect is clearly visible when the radius is larger than $\sim\SI{80}{nm}$.

\section{Conclusion}
Our results clearly show that the presence of a substrate can reduce or even suppress diffraction in particle arrays. This effect is important in experiments involving particles of small height relative to the period.\cite{AB08} Diffraction can be recovered in samples with larger metal particles.\cite{VGG09-2} These conclusions are important in the design of complex antennas involving interaction between metal parts at distances of several wavelengths on a substrate.\cite{BSK10} They may also offer ways of performing sensing by detecting small variations of index of refraction in a fluid environment.\cite{KSK10}
\section*{ACKNOWLEDGMENTS}
This work has been supported by the Spanish MICINN (MAT2007-66050 and Consolider NanoLight.es) and the European Commission (FP7-248909 "LIMA" and NMP4-SL-2008-213669 "ENSEMBLE"). XMB acknowledges support from a CSIC-JAE scholarship. WLB was a Wolfson Royal Society Merit Award holder.
\appendix
\section{Lattice sum for normal incidence}
\begin{figure}[!ht]
\centering
\centering \includegraphics[width=\columnwidth]{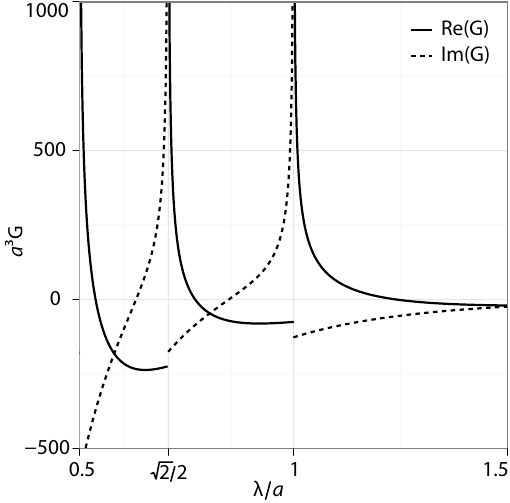}
\caption{Real and imaginary parts of the lattice sum $G$ as a function of wavelength $\lambda$ normalized to the period $a$ for normal incidence on a square array. The wavelength is given relative to the medium in which the lattice is immersed.}
\label{fig:lattice-sum}
\end{figure}

We represent in Fig.~\ref{fig:lattice-sum} the calculated lattice sum $G$, as defined in Eq.~(\ref{eqG}). This involves a poorly convergent sum that is accelerated by separating it in parts that are computed respectively in momentum and in real space, following the methods introduced by~\onlinecite{K1968}. The sum shows characteristic divergences at values of the wavelength for which a diffracted beam in the Rayleigh construction becomes grazing, as explained elsewhere.\cite{paper130}

\begin{thebibliography}{39}
\expandafter\ifx\csname natexlab\endcsname\relax\def\natexlab#1{#1}\fi
\expandafter\ifx\csname bibnamefont\endcsname\relax
  \def\bibnamefont#1{#1}\fi
\expandafter\ifx\csname bibfnamefont\endcsname\relax
  \def\bibfnamefont#1{#1}\fi
\expandafter\ifx\csname citenamefont\endcsname\relax
  \def\citenamefont#1{#1}\fi
\expandafter\ifx\csname url\endcsname\relax
  \def\url#1{\texttt{#1}}\fi
\expandafter\ifx\csname urlprefix\endcsname\relax\def\urlprefix{URL }\fi
\providecommand{\bibinfo}[2]{#2}
\providecommand{\eprint}[2][]{\url{#2}}

\bibitem[{\citenamefont{{Lord Rayleigh}}(1907{\natexlab{a}})}]{R1907-2}
\bibinfo{author}{\bibnamefont{{Lord Rayleigh}}}, \bibinfo{journal}{Proc.\ R.\
  Soc.\ Lond.\ A} \textbf{\bibinfo{volume}{79}}, \bibinfo{pages}{399}
  (\bibinfo{year}{1907}{\natexlab{a}}).

\bibitem[{\citenamefont{Wood}(1902)}]{W1902}
\bibinfo{author}{\bibfnamefont{R.~W.} \bibnamefont{Wood}},
  \bibinfo{journal}{Philos.\ Mag.} \textbf{\bibinfo{volume}{4}},
  \bibinfo{pages}{396} (\bibinfo{year}{1902}).

\bibitem[{\citenamefont{Fano}(1941)}]{F1941}
\bibinfo{author}{\bibfnamefont{U.}~\bibnamefont{Fano}}, \bibinfo{journal}{J.\
  Opt.\ Soc.\ Am.} \textbf{\bibinfo{volume}{31}}, \bibinfo{pages}{213}
  (\bibinfo{year}{1941}).

\bibitem[{\citenamefont{Zou and Schatz}(2004)}]{ZS04}
\bibinfo{author}{\bibfnamefont{S.}~\bibnamefont{Zou}} \bibnamefont{and}
  \bibinfo{author}{\bibfnamefont{G.~C.} \bibnamefont{Schatz}},
  \bibinfo{journal}{J.\ Chem.\ Phys.} \textbf{\bibinfo{volume}{121}},
  \bibinfo{pages}{12606} (\bibinfo{year}{2004}).

\bibitem[{\citenamefont{Markel and Sarychev}(2007)}]{MS07}
\bibinfo{author}{\bibfnamefont{V.~A.} \bibnamefont{Markel}} \bibnamefont{and}
  \bibinfo{author}{\bibfnamefont{A.~K.} \bibnamefont{Sarychev}},
  \bibinfo{journal}{Phys.\ Rev.\ B} \textbf{\bibinfo{volume}{75}},
  \bibinfo{pages}{085426} (\bibinfo{year}{2007}).

\bibitem[{\citenamefont{Markel and Sarychev}(2005)}]{M05-2}
\bibinfo{author}{\bibfnamefont{V.~A.} \bibnamefont{Markel}} \bibnamefont{and}
  \bibinfo{author}{\bibfnamefont{A.~K.} \bibnamefont{Sarychev}},
  \bibinfo{journal}{Phys.\ Rev.\ B} \textbf{\bibinfo{volume}{38}},
  \bibinfo{pages}{L115} (\bibinfo{year}{2005}).

\bibitem[{\citenamefont{Hicks et~al.}(2005)\citenamefont{Hicks, Zou, Schatz,
  Spears, {Van Duyne}, Gunnarsson, Rindzevicius, Kasemo, and K\"all}}]{HZS05}
\bibinfo{author}{\bibfnamefont{E.~M.} \bibnamefont{Hicks}},
  \bibinfo{author}{\bibfnamefont{S.}~\bibnamefont{Zou}},
  \bibinfo{author}{\bibfnamefont{G.~C.} \bibnamefont{Schatz}},
  \bibinfo{author}{\bibfnamefont{K.~G.} \bibnamefont{Spears}},
  \bibinfo{author}{\bibfnamefont{R.~P.} \bibnamefont{{Van Duyne}}},
  \bibinfo{author}{\bibfnamefont{L.}~\bibnamefont{Gunnarsson}},
  \bibinfo{author}{\bibfnamefont{T.}~\bibnamefont{Rindzevicius}},
  \bibinfo{author}{\bibfnamefont{B.}~\bibnamefont{Kasemo}}, \bibnamefont{and}
  \bibinfo{author}{\bibfnamefont{M.}~\bibnamefont{K\"all}},
  \bibinfo{journal}{Nano\ Lett.} \textbf{\bibinfo{volume}{5}},
  \bibinfo{pages}{1065} (\bibinfo{year}{2005}).

\bibitem[{\citenamefont{Linden et~al.}(2001)\citenamefont{Linden, Christ, Kuhl,
  and Giessen}}]{LCK01}
\bibinfo{author}{\bibfnamefont{S.}~\bibnamefont{Linden}},
  \bibinfo{author}{\bibfnamefont{A.}~\bibnamefont{Christ}},
  \bibinfo{author}{\bibfnamefont{J.}~\bibnamefont{Kuhl}}, \bibnamefont{and}
  \bibinfo{author}{\bibfnamefont{H.}~\bibnamefont{Giessen}},
  \bibinfo{journal}{Appl.\ Phys.\ B} \textbf{\bibinfo{volume}{73}},
  \bibinfo{pages}{311} (\bibinfo{year}{2001}).

\bibitem[{\citenamefont{Kravets et~al.}(2008)\citenamefont{Kravets, Schedin,
  and Grigorenko}}]{KSG08}
\bibinfo{author}{\bibfnamefont{V.~G.} \bibnamefont{Kravets}},
  \bibinfo{author}{\bibfnamefont{F.}~\bibnamefont{Schedin}}, \bibnamefont{and}
  \bibinfo{author}{\bibfnamefont{A.~N.} \bibnamefont{Grigorenko}},
  \bibinfo{journal}{Phys.\ Rev.\ Lett.} \textbf{\bibinfo{volume}{101}},
  \bibinfo{pages}{087403} (\bibinfo{year}{2008}).

\bibitem[{\citenamefont{Augui\'e and Barnes}(2008)}]{AB08}
\bibinfo{author}{\bibfnamefont{B.}~\bibnamefont{Augui\'e}} \bibnamefont{and}
  \bibinfo{author}{\bibfnamefont{W.~L.} \bibnamefont{Barnes}},
  \bibinfo{journal}{Phys.\ Rev.\ Lett.} \textbf{\bibinfo{volume}{101}},
  \bibinfo{pages}{143902} (\bibinfo{year}{2008}).

\bibitem[{\citenamefont{Chu et~al.}(2008)\citenamefont{Chu, Schonbrun, Yang,
  and Crozier}}]{CSY08}
\bibinfo{author}{\bibfnamefont{Y.}~\bibnamefont{Chu}},
  \bibinfo{author}{\bibfnamefont{E.}~\bibnamefont{Schonbrun}},
  \bibinfo{author}{\bibfnamefont{T.}~\bibnamefont{Yang}}, \bibnamefont{and}
  \bibinfo{author}{\bibfnamefont{K.~B.} \bibnamefont{Crozier}},
  \bibinfo{journal}{Appl.\ Phys.\ Lett.} \textbf{\bibinfo{volume}{93}},
  \bibinfo{pages}{181108} (\bibinfo{year}{2008}).

\bibitem[{\citenamefont{Vecchi et~al.}(2009{\natexlab{a}})\citenamefont{Vecchi,
  Giannini, and {G\'omez Rivas}}}]{VGG09}
\bibinfo{author}{\bibfnamefont{G.}~\bibnamefont{Vecchi}},
  \bibinfo{author}{\bibfnamefont{V.}~\bibnamefont{Giannini}}, \bibnamefont{and}
  \bibinfo{author}{\bibfnamefont{J.}~\bibnamefont{{G\'omez Rivas}}},
  \bibinfo{journal}{Phys.\ Rev.\ B} \textbf{\bibinfo{volume}{80}},
  \bibinfo{pages}{201401(R)} (\bibinfo{year}{2009}{\natexlab{a}}).

\bibitem[{\citenamefont{Novotny}(2007)}]{N07_2}
\bibinfo{author}{\bibfnamefont{L.}~\bibnamefont{Novotny}},
  \bibinfo{journal}{Phys.\ Rev.\ Lett.} \textbf{\bibinfo{volume}{98}},
  \bibinfo{pages}{266802} (\bibinfo{year}{2007}).

\bibitem[{\citenamefont{Maier et~al.}(2003)\citenamefont{Maier, Kik, Atwater,
  Meltzer, Harel, Koel, and Requicha}}]{MKA03}
\bibinfo{author}{\bibfnamefont{S.~A.} \bibnamefont{Maier}},
  \bibinfo{author}{\bibfnamefont{P.~G.} \bibnamefont{Kik}},
  \bibinfo{author}{\bibfnamefont{H.~A.} \bibnamefont{Atwater}},
  \bibinfo{author}{\bibfnamefont{S.}~\bibnamefont{Meltzer}},
  \bibinfo{author}{\bibfnamefont{E.}~\bibnamefont{Harel}},
  \bibinfo{author}{\bibfnamefont{B.~E.} \bibnamefont{Koel}}, \bibnamefont{and}
  \bibinfo{author}{\bibfnamefont{A.~A.~G.} \bibnamefont{Requicha}},
  \bibinfo{journal}{Nat.\ Mater.} \textbf{\bibinfo{volume}{2}},
  \bibinfo{pages}{229} (\bibinfo{year}{2003}).

\bibitem[{\citenamefont{Sainidou and {Garc\'{\i}a de Abajo}}(2008)}]{paper148}
\bibinfo{author}{\bibfnamefont{R.}~\bibnamefont{Sainidou}} \bibnamefont{and}
  \bibinfo{author}{\bibfnamefont{F.~J.} \bibnamefont{{Garc\'{\i}a de Abajo}}},
  \bibinfo{journal}{Opt.\ Express} \textbf{\bibinfo{volume}{16}},
  \bibinfo{pages}{4499} (\bibinfo{year}{2008}).

\bibitem[{\citenamefont{Enoch et~al.}(2004)\citenamefont{Enoch, Quidant, and
  Badenes}}]{EQB04}
\bibinfo{author}{\bibfnamefont{S.}~\bibnamefont{Enoch}},
  \bibinfo{author}{\bibfnamefont{R.}~\bibnamefont{Quidant}}, \bibnamefont{and}
  \bibinfo{author}{\bibfnamefont{G.}~\bibnamefont{Badenes}},
  \bibinfo{journal}{Opt.\ Express} \textbf{\bibinfo{volume}{29}},
  \bibinfo{pages}{3422} (\bibinfo{year}{2004}).

\bibitem[{\citenamefont{Kravets et~al.}(2010)\citenamefont{Kravets, Schedin,
  Kabashin, and Grigorenko}}]{KSK10}
\bibinfo{author}{\bibfnamefont{V.~G.} \bibnamefont{Kravets}},
  \bibinfo{author}{\bibfnamefont{F.}~\bibnamefont{Schedin}},
  \bibinfo{author}{\bibfnamefont{A.~V.} \bibnamefont{Kabashin}},
  \bibnamefont{and} \bibinfo{author}{\bibfnamefont{A.~N.}
  \bibnamefont{Grigorenko}}, \bibinfo{journal}{Opt. Lett.}
  \textbf{\bibinfo{volume}{35}}, \bibinfo{pages}{956} (\bibinfo{year}{2010}).

\bibitem[{\citenamefont{Vecchi et~al.}(2009{\natexlab{b}})\citenamefont{Vecchi,
  Giannini, and {G\'omez Rivas}}}]{VGG09-2}
\bibinfo{author}{\bibfnamefont{G.}~\bibnamefont{Vecchi}},
  \bibinfo{author}{\bibfnamefont{V.}~\bibnamefont{Giannini}}, \bibnamefont{and}
  \bibinfo{author}{\bibfnamefont{J.}~\bibnamefont{{G\'omez Rivas}}},
  \bibinfo{journal}{Phys.\ Rev.\ Lett.} \textbf{\bibinfo{volume}{102}},
  \bibinfo{pages}{146807} (\bibinfo{year}{2009}{\natexlab{b}}).

\bibitem[{\citenamefont{Curto et~al.}(2010)\citenamefont{Curto, Volpe,
  Taminiau, Kreuzer, Quidant, and van Hulst}}]{CVT10}
\bibinfo{author}{\bibfnamefont{A.~G.} \bibnamefont{Curto}},
  \bibinfo{author}{\bibfnamefont{G.}~\bibnamefont{Volpe}},
  \bibinfo{author}{\bibfnamefont{T.~H.} \bibnamefont{Taminiau}},
  \bibinfo{author}{\bibfnamefont{M.~P.} \bibnamefont{Kreuzer}},
  \bibinfo{author}{\bibfnamefont{R.}~\bibnamefont{Quidant}}, \bibnamefont{and}
  \bibinfo{author}{\bibfnamefont{N.~F.} \bibnamefont{van Hulst}},
  \bibinfo{journal}{Science} \textbf{\bibinfo{volume}{329}}
  (\bibinfo{year}{2010}).

\bibitem[{\citenamefont{Augui\'e and Barnes}(2009)}]{AB09}
\bibinfo{author}{\bibfnamefont{B.}~\bibnamefont{Augui\'e}} \bibnamefont{and}
  \bibinfo{author}{\bibfnamefont{W.~L.} \bibnamefont{Barnes}},
  \bibinfo{journal}{Opt.\ Lett.} \textbf{\bibinfo{volume}{34}},
  \bibinfo{pages}{401} (\bibinfo{year}{2009}).

\bibitem[{\citenamefont{Grzelczak et~al.}(2008)\citenamefont{Grzelczak,
  {P\'erez-Juste}, Mulvaney, , and {Liz-Marz\'an}}}]{GPM08}
\bibinfo{author}{\bibfnamefont{M.}~\bibnamefont{Grzelczak}},
  \bibinfo{author}{\bibfnamefont{J.}~\bibnamefont{{P\'erez-Juste}}},
  \bibinfo{author}{\bibfnamefont{P.}~\bibnamefont{Mulvaney}}, ,
  \bibnamefont{and} \bibinfo{author}{\bibfnamefont{L.~M.}
  \bibnamefont{{Liz-Marz\'an}}}, \bibinfo{journal}{Chem.\ Soc.\ Rev.}
  \textbf{\bibinfo{volume}{37}}, \bibinfo{pages}{1783} (\bibinfo{year}{2008}).

\bibitem[{\citenamefont{{Benda\~{n}a} et~al.}(2009)\citenamefont{{Benda\~{n}a},
  {Garc\'{\i}a de Abajo}, and Polman}}]{paper186}
\bibinfo{author}{\bibfnamefont{X.~M.} \bibnamefont{{Benda\~{n}a}}},
  \bibinfo{author}{\bibfnamefont{F.~J.} \bibnamefont{{Garc\'{\i}a de Abajo}}},
  \bibnamefont{and} \bibinfo{author}{\bibfnamefont{A.}~\bibnamefont{Polman}},
  \bibinfo{journal}{Opt.\ Express} \textbf{\bibinfo{volume}{17}},
  \bibinfo{pages}{18826} (\bibinfo{year}{2009}).

\bibitem[{\citenamefont{Stefanou et~al.}(1992)\citenamefont{Stefanou,
  Karathanos, and Modinos}}]{SKM92}
\bibinfo{author}{\bibfnamefont{N.}~\bibnamefont{Stefanou}},
  \bibinfo{author}{\bibfnamefont{V.}~\bibnamefont{Karathanos}},
  \bibnamefont{and} \bibinfo{author}{\bibfnamefont{A.}~\bibnamefont{Modinos}},
  \bibinfo{journal}{J.\ Phys.\ Condens.\ Matter} \textbf{\bibinfo{volume}{4}},
  \bibinfo{pages}{7389} (\bibinfo{year}{1992}).

\bibitem[{\citenamefont{Johnson and Christy}(1972)}]{JC1972}
\bibinfo{author}{\bibfnamefont{P.~B.} \bibnamefont{Johnson}} \bibnamefont{and}
  \bibinfo{author}{\bibfnamefont{R.~W.} \bibnamefont{Christy}},
  \bibinfo{journal}{Phys.\ Rev.\ B} \textbf{\bibinfo{volume}{6}},
  \bibinfo{pages}{4370} (\bibinfo{year}{1972}).

\bibitem[{\citenamefont{Yamaguchi et~al.}(1974)\citenamefont{Yamaguchi,
  Yoshida, and Kinbara}}]{YYK74}
\bibinfo{author}{\bibfnamefont{T.}~\bibnamefont{Yamaguchi}},
  \bibinfo{author}{\bibfnamefont{S.}~\bibnamefont{Yoshida}}, \bibnamefont{and}
  \bibinfo{author}{\bibfnamefont{A.}~\bibnamefont{Kinbara}},
  \bibinfo{journal}{Thin Solid Films} \textbf{\bibinfo{volume}{21}},
  \bibinfo{pages}{173 } (\bibinfo{year}{1974}), ISSN \bibinfo{issn}{0040-6090}.

\bibitem[{\citenamefont{Meier et~al.}(1985)\citenamefont{Meier, Wokaun, and
  Liao}}]{MWL1985}
\bibinfo{author}{\bibfnamefont{M.}~\bibnamefont{Meier}},
  \bibinfo{author}{\bibfnamefont{A.}~\bibnamefont{Wokaun}}, \bibnamefont{and}
  \bibinfo{author}{\bibfnamefont{P.~F.} \bibnamefont{Liao}},
  \bibinfo{journal}{J.\ Opt.\ Soc.\ Am.\ B} \textbf{\bibinfo{volume}{2}},
  \bibinfo{pages}{931} (\bibinfo{year}{1985}).

\bibitem[{\citenamefont{Johansson}(2001)}]{J01}
\bibinfo{author}{\bibfnamefont{P.}~\bibnamefont{Johansson}},
  \bibinfo{journal}{Phys.\ Rev.\ B} \textbf{\bibinfo{volume}{64}},
  \bibinfo{pages}{165405} (\bibinfo{year}{2001}).

\bibitem[{\citenamefont{Stefanou and Modinos}(1991)}]{SM91}
\bibinfo{author}{\bibfnamefont{N.}~\bibnamefont{Stefanou}} \bibnamefont{and}
  \bibinfo{author}{\bibfnamefont{A.}~\bibnamefont{Modinos}},
  \bibinfo{journal}{J.\ Phys.\ Condens.\ Matter} \textbf{\bibinfo{volume}{3}},
  \bibinfo{pages}{8135} (\bibinfo{year}{1991}).

\bibitem[{\citenamefont{Soller and Hall}(2002)}]{SH02}
\bibinfo{author}{\bibfnamefont{B.~J.} \bibnamefont{Soller}} \bibnamefont{and}
  \bibinfo{author}{\bibfnamefont{D.~G.} \bibnamefont{Hall}},
  \bibinfo{journal}{J.\ Opt.\ Soc.\ Am.\ B} \textbf{\bibinfo{volume}{19}},
  \bibinfo{pages}{2437} (\bibinfo{year}{2002}).

\bibitem[{\citenamefont{Simsek}(2009)}]{S09}
\bibinfo{author}{\bibfnamefont{E.}~\bibnamefont{Simsek}},
  \bibinfo{journal}{Plasmonics} \textbf{\bibinfo{volume}{4}},
  \bibinfo{pages}{223} (\bibinfo{year}{2009}).

\bibitem[{\citenamefont{Simsek}(2010)}]{S10}
\bibinfo{author}{\bibfnamefont{E.}~\bibnamefont{Simsek}},
  \bibinfo{journal}{Opt.\ Express} \textbf{\bibinfo{volume}{18}},
  \bibinfo{pages}{1722} (\bibinfo{year}{2010}).

\bibitem[{\citenamefont{Vernon et~al.}(2010)\citenamefont{Vernon, Funston,
  Novo, G{\'o}mez, Mulvaney, and Davis}}]{VFN10}
\bibinfo{author}{\bibfnamefont{K.~C.} \bibnamefont{Vernon}},
  \bibinfo{author}{\bibfnamefont{A.~M.} \bibnamefont{Funston}},
  \bibinfo{author}{\bibfnamefont{C.}~\bibnamefont{Novo}},
  \bibinfo{author}{\bibfnamefont{D.~E.} \bibnamefont{G{\'o}mez}},
  \bibinfo{author}{\bibfnamefont{P.}~\bibnamefont{Mulvaney}}, \bibnamefont{and}
  \bibinfo{author}{\bibfnamefont{E.}~\bibnamefont{Davis}},
  \bibinfo{journal}{Nano\ Lett.}  (\bibinfo{year}{2010}).

\bibitem[{\citenamefont{{Kuwata-Gonokami}
  et~al.}(2003)\citenamefont{{Kuwata-Gonokami}, Tamaru, Esumi, and
  Miyano}}]{KTE03}
\bibinfo{author}{\bibfnamefont{H.}~\bibnamefont{{Kuwata-Gonokami}}},
  \bibinfo{author}{\bibfnamefont{H.}~\bibnamefont{Tamaru}},
  \bibinfo{author}{\bibfnamefont{K.}~\bibnamefont{Esumi}}, \bibnamefont{and}
  \bibinfo{author}{\bibfnamefont{K.}~\bibnamefont{Miyano}},
  \bibinfo{journal}{Appl.\ Phys.\ Lett.} \textbf{\bibinfo{volume}{83}},
  \bibinfo{pages}{4625} (\bibinfo{year}{2003}).

\bibitem[{\citenamefont{{Garc\'{\i}a de Abajo}}(2007)}]{paper130}
\bibinfo{author}{\bibfnamefont{F.~J.} \bibnamefont{{Garc\'{\i}a de Abajo}}},
  \bibinfo{journal}{Rev.\ Mod.\ Phys.} \textbf{\bibinfo{volume}{79}},
  \bibinfo{pages}{1267} (\bibinfo{year}{2007}).

\bibitem[{\citenamefont{Hessel and Oliner}(1965)}]{HO65}
\bibinfo{author}{\bibfnamefont{A.}~\bibnamefont{Hessel}} \bibnamefont{and}
  \bibinfo{author}{\bibfnamefont{A.~A.} \bibnamefont{Oliner}},
  \bibinfo{journal}{Appl. Opt.} \textbf{\bibinfo{volume}{4}},
  \bibinfo{pages}{1275} (\bibinfo{year}{1965}).

\bibitem[{\citenamefont{Doicu et~al.}(2006)\citenamefont{Doicu, Wriedt, and
  Eremin}}]{DWE06}
\bibinfo{author}{\bibfnamefont{A.}~\bibnamefont{Doicu}},
  \bibinfo{author}{\bibfnamefont{T.}~\bibnamefont{Wriedt}}, \bibnamefont{and}
  \bibinfo{author}{\bibfnamefont{Y.~A.} \bibnamefont{Eremin}},
  \emph{\bibinfo{title}{Light Scattering by Systems of Particles: Null-Field
  Method with Discrete Sources: Theory and Programs}}
  (\bibinfo{publisher}{Springer-Verlag}, \bibinfo{address}{Berlin},
  \bibinfo{year}{2006}).

\bibitem[{\citenamefont{{Lord Rayleigh}}(1907{\natexlab{b}})}]{R1907}
\bibinfo{author}{\bibnamefont{{Lord Rayleigh}}}, \bibinfo{journal}{Philos.\
  Mag.} \textbf{\bibinfo{volume}{14}}, \bibinfo{pages}{60}
  (\bibinfo{year}{1907}{\natexlab{b}}).

\bibitem[{\citenamefont{Brinks et~al.}(2010)\citenamefont{Brinks, Stefani,
  Kulzer, Hildner, Taminiau, Avlasevich, M\"ullen, and {van Hulst}}}]{BSK10}
\bibinfo{author}{\bibfnamefont{D.}~\bibnamefont{Brinks}},
  \bibinfo{author}{\bibfnamefont{F.~D.} \bibnamefont{Stefani}},
  \bibinfo{author}{\bibfnamefont{F.}~\bibnamefont{Kulzer}},
  \bibinfo{author}{\bibfnamefont{R.}~\bibnamefont{Hildner}},
  \bibinfo{author}{\bibfnamefont{T.~H.} \bibnamefont{Taminiau}},
  \bibinfo{author}{\bibfnamefont{Y.}~\bibnamefont{Avlasevich}},
  \bibinfo{author}{\bibfnamefont{K.}~\bibnamefont{M\"ullen}}, \bibnamefont{and}
  \bibinfo{author}{\bibfnamefont{N.~F.} \bibnamefont{{van Hulst}}},
  \bibinfo{journal}{Nature} \textbf{\bibinfo{volume}{465}},
  \bibinfo{pages}{905} (\bibinfo{year}{2010}).

\bibitem[{\citenamefont{Kambe}(1968)}]{K1968}
\bibinfo{author}{\bibfnamefont{K.}~\bibnamefont{Kambe}}, \bibinfo{journal}{Z.\
  Naturforsch.\ A} \textbf{\bibinfo{volume}{23}}, \bibinfo{pages}{1280}
  (\bibinfo{year}{1968}).

\end{thebibliography}

\end{document}